\documentstyle[preprint,prl,aps]{revtex}

\begin{document}

\draft
\title{
  \begin{flushright}
    \large UT-743
  \end{flushright}
  Natural Unification with a Supersymmetric SO(10)$_{GUT} \:\times$
  SO(6)$_H$ Gauge Theory}
\author{T. Hotta, Izawa K.-I. and T. Yanagida}
\address{Department of Physics, University of Tokyo, Bunkyo-ku, Tokyo
  113, Japan}
\date{\today}
\maketitle

\begin{abstract}
We propose a unified model of elementary particles
based on a supersymmetric SO(10)$_{GUT} \:\times$ SO(6)$_H$ gauge theory.
This model completely achieves natural unification of the strong
and electroweak interactions without any fine-tunings.
\end{abstract}


It has been long noticed that the quarks and leptons fit into complete 
multiplets of a simple gauge group such as SU(5)$_{GUT}$ \cite{GUT}
and SO(10)$_{GUT}$  \cite{SOten} which unifies the standard gauge
group.
This grand unification (GUT) approach seems to yield an attractive
candidate of a unified model beyond the standard one.
In fact, recent high-precision measurements of the three gauge
coupling constants at low energies strongly support reality of the
supersymmetric (SUSY) unification of this kind \cite{preciceexp}.
However, as for the Higgs doublet, which also takes a crucial part in
the standard model, the unification does not work naturally. 
The doublet tends to be accompanied by its triplet partner due to
simplicity of the gauge group, which poses a fine-tuning problem of
the doublet-triplet splitting \cite{dtsplit}.

In this letter, we propose a supersymmetric unified model based on a
semisimple gauge group SO(10)$_{GUT} \:\times$ SO(6)$_H$, and show
that it completely achieves natural unification of elementary
particles without any fine-tunings \cite{ours}, keeping all the
excellent successes of minimal SO(10) SUSY-GUT's.
In particular, the doublet-triplet splitting in the Higgs sector is
realized naturally in this model.
At first glance, non-simplicity of the gauge group appears to spoil
unification of the standard three gauge coupling constants, which
would be achieved by means of a simple gauge group.
However, that is not the case.
It suffices that the SO(6)$_H$ gauge interaction is strong enough at
the GUT scale \cite{our1,our2}, which needs no fine-tuning (in a gauge
theory with asymptotic freedom).

Let us first consider a SUSY SO(10)$_{GUT} \:\times$ SO($N_C$)$_H$
gauge theory with eleven flavors of hyperquarks $Q_\alpha^A \ (A = 1,
\cdots, 11; \ \alpha = 1, \cdots, N_C)$ which transform as the vector
{\boldmath $N_C$} representation under the hypercolor SO($N_C$)$_H$.
The first ten hyperquarks $Q_\alpha^I \ (I = 1, \cdots, 10)$ form the
{\bf 10} representation and the last one $Q^{11}_\alpha$ is a singlet
of the SO(10)$_{GUT}$.
The reason why we take the SO($N_C$)$_H$ (rather than the
SU($N_C$)$_H$) is that the anomaly-free condition is automatically
satisfied for the orthogonal groups (except SO(6) $\simeq$ SU(4))
\cite{anomaly}.
We also introduce two kinds of SO($N_C$)$_H$-singlet chiral
superfields --- a {\bf 10} Higgs $H_I$ and a {\bf 54} Higgs $S_{IJ} \
(I, J = 1, \cdots, 10)$ of the SO(10)$_{GUT}$.

All the quark and lepton superfields constitute spinor {\bf 16}
representations of the SO(10)$_{GUT}$ and singlets of the
SO($N_C$)$_H$.
We choose $N_C \geq 6$ to guarantee the asymptotic freedom of the
hypercolor gauge interaction.
In the following, we restrict ourselves to the minimum case $N_C = 6$.

Imposing a global U(1)$_A$ symmetry
\begin{equation}
  \label{uoneA}
  Q_\alpha^I \rightarrow Q_\alpha^I , \
  Q_\alpha^{11} \rightarrow e^{-2 i \xi} Q_\alpha^{11} , \
  H_I \rightarrow e^{2 i \xi} H_I , \
  S_{IJ} \rightarrow S_{IJ} , 
\end{equation}
we forbid such (mass) terms as $H_I H_I$ and $Q_\alpha^{11}
Q_\alpha^{11}$ in the superpotential.
Notice that this global U(1)$_A$ has an anomaly --- it is broken by
quantum effects.
However, the broken U(1)$_A$ still plays a crucial role to guarantee
the masslessness of Higgs doublets \cite{our2}.

The tree-level superpotential is, then, given by
\begin{equation}
  \label{potQ}
  W = \lambda_Q \, Q^I_\alpha Q^J_\alpha S_{IJ}
  + m_Q \, Q^I_\alpha Q^I_\alpha + h Q_\alpha^I Q_\alpha^{11} H_I
  + \frac{1}{2} m_S Tr(S^2) + \frac{1}{3} \lambda_S Tr (S^3) , \ \ \
  (I, J = 1, \cdots, 10) .
\end{equation}
Classically there is an SO(10)$_{GUT}$-unbroken vacuum, whereas it
does not exist quantum mechanically.
If $\langle S_{IJ} \rangle = 0$, the low-energy physics below the
scale $m_Q \neq 0$ would be effectively described by an SO(6)$_H$
gauge theory with one massless hyperquark $Q^{11}_\alpha$. For this
case $(N_f \leq N_C - 5)$, however, there is no stable SUSY vacuum
\cite{dualitySO}.
Here, $N_f$ denotes the number of flavors of massless hyperquarks.

We now show that there is indeed a desired SUSY vacuum in the present
model.
We consider vacua specified by
\newfont{\bg}{cmbx10 scaled\magstep2}
\begin{equation}
  \label{vacuumS}
  \langle S_{IJ} \rangle = v
  \left(
    \begin{array}{ccccc}
      {3 \over 2} \smash{\lower0.2ex\hbox{\bg 1}} & & & & \\
      & {3 \over 2} \smash{\lower0.2ex\hbox{\bg 1}} & & & \\
      & & - \smash{\lower0.2ex\hbox{\bg 1}} & & \\
      & & & - \smash{\lower0.2ex\hbox{\bg 1}} & \\
      & & & & - \smash{\lower0.2ex\hbox{\bg 1}} 
    \end{array}
  \right); \quad
  \smash{\lower0.2ex\hbox{\bg 1}} = \left (
  \begin{array}{cc}
    1 & 0\\
    0 & 1
  \end{array} \right ) ,
\end{equation}
in which the SO(10)$_{GUT}$ is broken down to the Pati-Salam gauge
group SO(6) $\times$ SU(2)$_L \:\times$ SU(2)$_R$ \cite{Pati}.
The vacuum-expectation value $v$ is dynamically chosen as $\lambda_Q v 
= m_Q$ or $-\frac{2}{3} m_Q$ so that there are seven or five
massless hyperquarks, respectively \cite{vacuum}.
We adopt the former vacuum, since the latter has a phenomenological 
problem \cite{anothervacuum}.

In the vacua $v = m_Q / \lambda_Q$, the first four hyperquarks
$Q_\alpha^i \ (i = 1, \cdots, 4)$ are massive and the rest
$Q_\alpha^a \ (a = 5, \cdots, 11)$ are massless.
Therefore, we may treat this theory as an SO(6)$_H$ gauge theory with
seven hyperquarks at low energies.
Fortunately, the quantum moduli space is the same as the classical one
for $N_f \geq  N_C - 1$ in the SO($N_C$)$_H$ gauge theory
\cite{dualitySO}.
This allows us to analyze the vacua with the aid of the classical
theory in Eq.(\ref{potQ}), since the numbers of flavors and colors for
the massless hyperquarks $Q_\alpha^a$ amount to $N_f = 7$ and $N_C =
6$, respectively.

It is now a straightforward task to seek all of the vacua with
Eq.(\ref{vacuumS}).
The $F$-term flatness conditions for the vacua are given by
\begin{equation}
  \begin{array}{l}
    \displaystyle
    m_S S^{IJ} + \lambda_S \left( (S^2)^{IJ} - \frac{1}{10} Tr(S^2)
    \delta^{IJ} \right)
    + \lambda_Q \left( Q^I_\alpha Q^J_\alpha - \frac{1}{10}
    Tr(Q^I_\alpha Q^J_\alpha) \delta^{IJ} \right) = 0 , \\
    \noalign{\vskip 0.5ex}
    2 ( \lambda_Q \, S_{IJ} + m_Q \delta_{IJ}) Q^J_\alpha
    + h Q^{11}_\alpha H_I = 0 , \\ 
    \noalign{\vskip 1ex}
    h Q_\alpha^I H_I = 0 , \\
    \noalign{\vskip 1ex}
    h Q_\alpha^I Q_\alpha^{11} = 0 .
  \end{array}
\end{equation}
From these equations together with the $D$-term flatness conditions,
we find the following vacuum:
\newfont{\bgg}{cmbx10 scaled\magstep3}
\begin{equation}
  \label{vacuum}
  \langle Q_\alpha^I \rangle = v_Q
  \left(
    \begin{array}{ccc}
      & \smash{\lower1.2ex\hbox{\bgg O}} & \\
      \noalign{\vskip 1ex}
      \smash{\lower0.2ex\hbox{\bg 1}} & & \\
      & \smash{\lower0.2ex\hbox{\bg 1}} & \\
      & & \smash{\lower0.2ex\hbox{\bg 1}}
    \end{array}
  \right), \quad
  \langle Q_\alpha^{11} \rangle = 0, \quad
  \langle H_I \rangle = 0 ; \ \ 
  v_Q = \sqrt{\frac{5 (2 \lambda_Q m_S \, m_Q + \lambda_S m_Q^2)}{4
      \lambda_Q{}^3}} ,
\end{equation}
with Eq.(\ref{vacuumS}).
We take the Yukawa couplings $\lambda_Q \sim \lambda_S \sim {\cal
  O}(1)$ and $v \sim v_Q$, assuming $m_S \sim m_Q$.
The GUT scale $v \sim v_Q$ is suggested to be $\sim 10^{16}$ GeV from
the renormalization group analysis on the gauge coupling constants of
the low-energy gauge groups \cite{preciceexp}. 

In this vacuum, the $Q_\alpha^I$ condensation in Eq.(\ref{vacuum})
breaks the hypercolor SO(6)$_H$ gauge symmetry (i.e. a Higgs phase is
realized) and the total gauge group is broken down to the SO(6)$_C
\:\times$ SU(2)$_L \:\times$ SU(2)$_R$.
Here, the unbroken SO(6)$_C$ is the diagonal subgroup of the
hypercolor SO(6)$_H$ and the SO(6) subgroup of SO(10)$_{GUT}$. The
quark-lepton superfields transform as ({\bf 4, 2, 1}) and ({\bf 4$^*$,
  1, 2}) under this unbroken SO(6)$_C \:\times$ SU(2)$_L \:\times$
SU(2)$_R$, as they do in the original Pati-Salam model.
The gauge coupling constant of the SO(6)$_C$ is given by
\begin{equation}
  \alpha_6 = \frac{\alpha_{GUT}}{1 + \alpha_{GUT} / \alpha_H}.
\end{equation}
In general $\alpha_6 \neq \alpha_{GUT}$ and hence we do not have the
GUT unification of the gauge coupling constants of the low-energy
gauge groups.
However, as stressed before, the GUT unification is achieved naturally
in a strong coupling limit of the hypercolor SO(6)$_H$, i.e. $\alpha_H
\gg 1$ \cite{coupling}. 

Since the masses $m_Q$ and $m_S$ in the superpotential in
Eq.(\ref{potQ}) are assumed to be of order the GUT scale, the
vacuum-expectation value $v_Q$ of $Q_\alpha^a \ (a = 5, \cdots, 10)$
also takes a value of the GUT scale.
In view of Eq.(\ref{potQ}), the colored Higgs $H_a \ (a = 5, \cdots,
10)$ obtain the GUT scale masses together with $Q_\alpha^{11}$ but the
Higgs $H_i \ (i = 1, \cdots, 4)$ remain massless as long as $\langle
Q_\alpha^{11} \rangle = 0$.
These massless Higgs multiplets transform as ({\bf 2}, {\bf 2}) under
the SU(2)$_L \:\times$ SU(2)$_R$.
They are identified with two Higgs doublets in the SUSY standard model
when the SO(6)$_C \:\times$ SU(2)$_L \:\times$ SU(2)$_R$ is broken
down to the SU(3)$_C \:\times$ SU(2)$_L \:\times$ U(1)$_Y$.
The masslessness of $H_i$ is guaranteed by the U(1)$_A$ symmetry in
Eq.(\ref{uoneA}) \cite{our2}.
On the other hand, the mass term for the colored Higgs $H_a$ and
$Q_\alpha^{11}$ is allowed since they have the opposite U(1)$_A$
charges each other.
Notice that the presence of the vacuum with unbroken U(1)$_A$ in
Eqs.(\ref{vacuumS}) and (\ref{vacuum}) is a dynamical consequence of the
present model.

We are now at a point to discuss further breaking of the remaining
SO(6)$_C \:\times$ SU(2)$_L \:\times$ SU(2)$_R$ down to the SU(3)$_C 
\:\times$ SU(2)$_L \:\times$ U(1)$_Y$.
To this end, we introduce additional Higgs superfields $A_{IJ}(\bf
45)$, $\phi(\bf 16)$, $\bar{\phi}(\bf 16^*)$ and a singlet $\chi$ of
the SO(10)$_{GUT}$ \cite{chi}.
Their superpotential is given by
\begin{equation}
  \label{potA}
  W' = m_A Tr(A^2) + \lambda_A Tr(A^2 S)
  + g_\phi (\bar{\phi} \sigma_{IJ} \phi) A_{IJ}
  + g_\chi (\bar{\phi} \phi - \mu^2) \chi.
\end{equation}
Then, the total superpotential is given by the sum of Eq.(\ref{potQ})
and Eq.(\ref{potA}).

In order to show the presence of a desired vacuum,
we consider the following
forms of $\langle A_{IJ} \rangle$ and $\langle \phi \rangle$:
\begin{equation}
  \label{vacuumA}
  \langle A_{IJ} \rangle =
  \left(
    \begin{array}{ccccc}
      a \, i \sigma_2 & & & & \\
      & a \, i \sigma_2 & & & \\
      & & b \, i \sigma_2 & & \\
      & & & b \, i \sigma_2 & \\
      & & & & b \, i \sigma_2
    \end{array}
  \right), \
  \langle \phi \rangle = v_\phi \left(\uparrow \otimes \uparrow \otimes
  \uparrow \otimes \uparrow \otimes \uparrow \right), 
\end{equation}
where \cite{notation}
\begin{equation}
  i\sigma_2 = \left (
  \begin{array}{cc}
    0 & 1 \\
    -1 & 0
  \end{array} \right ).
\end{equation}
The vacuum-expectation values $a, b, v_\phi, v_Q$ and $\chi$ satisfy
equations
\begin{equation}
  \begin{array}{l}
    \displaystyle - 5 m_Q m_S \lambda_Q^{-1}
    + 2 \lambda_A (a^2 - b^2) 
    - \frac{5}{2} \lambda_S m_Q^2 \lambda_Q^{-2} + 2 \lambda_Q v_Q^2
    = 0, \\
    \noalign{\vskip 1ex}
    \displaystyle
    C_1 a - g_\phi v_\phi^2 = 0, \\
    \noalign{\vskip 1ex}
    \displaystyle
    C_2 b - g_\phi v_\phi^2 = 0, \\
    \noalign{\vskip 1ex}
    g_\phi (4 a + 6 b) v_\phi + g_\chi v_\phi \chi = 0, \\
    \noalign{\vskip 0.5ex}
    g_\chi (v_\phi^2 - \mu^2) = 0, 
  \end{array}
\end{equation}
where we define
\begin{equation}
  \displaystyle C_1 \equiv 2 m_A + 3 \lambda_A m_Q \lambda_Q^{-1}, \quad
  \displaystyle C_2 \equiv 2 m_A - 2 \lambda_A m_Q \lambda_Q^{-1}.
\end{equation}
Here, we have used $v = m_Q / \lambda_Q$ in Eq.(\ref{vacuumS}).
From these equations we find a vacuum
\begin{equation}
  \label{vacuumAll}
  \begin{array}{l}
    \displaystyle 
    v_Q = \frac{5}{2} m_Q m_S \lambda_Q^{-2}
    + \frac{5}{4} \lambda_S m_Q^2 \lambda_Q^{-3}
    - g_\phi^2 \mu^4 \lambda_A \lambda_Q^{-1} (C_1^2 - C_2^2)^{-1}, \\ 
    \noalign{\vskip 1ex}
    \displaystyle
    a = g_\phi \mu^2 C_1^{-1}, \\
    \noalign{\vskip 1ex}
    \displaystyle
    b = g_\phi \mu^2 C_2^{-1}, \\
    \noalign{\vskip 1ex}
    \chi = - \displaystyle 10 g_\phi^2 \mu^2 
    \left( \lambda_A m_Q \lambda_Q^{-1} + 2 m_A
    \right) \displaystyle (g_\chi C_1 C_2)^{-1}, \\
    \noalign{\vskip 0.5ex}
    v_\phi^2 = \mu^2,
  \end{array}
\end{equation}
where the total gauge group SO(10)$_{GUT} \:\times$ SO(6)$_H$ is now
broken down to just the standard gauge group SU(3)$_C \:\times$
SU(2)$_L \:\times$ U(1)$_Y$ at the GUT scale.

It is important that the $\langle Q_\alpha^{11} \rangle$ still
vanishes as it does in Eq.(\ref{vacuum}).
Thus, the masslessness of the Higgs doublets $H_i$ is maintained in
this final vacuum.

Now, we proceed to the ordinary matter sector.
We assign one unit of the U(1)$_A$ charge to the quark-lepton
superfields so that the Higgs doublets $H_i$ couple to them and
produce their masses after spontaneous breaking of the SU(2)$_L
\:\times$ U(1)$_Y$ symmetry.
Then, there are two issues about quark and lepton masses as in the
minimal SO(10)$_{GUT}$ models: vanishing mixing angles in the quark
sector and mass degeneracy of the neutrinos and up-type quarks.

These issues are resolved by taking into account nonrenormalizable
interactions.
Let us define the U(1)$_A$ transformation of $\phi$ and $\bar{\phi}$
as
\begin{equation}
  \phi \rightarrow e^{- i \xi} \phi , \ 
  \bar{\phi} \rightarrow e^{i \xi} \bar{\phi} .
\end{equation}
Then, the superpotential of the model may contain nonrenormalizable
interactions
\begin{equation}
  W'' = \frac{f_{\alpha \beta}}{M_P} \psi_\alpha \psi_\beta
  \bar{\phi} \bar{\phi},
\end{equation}
where $M_P$ denotes the Planck scale as a cutoff and $\psi_\alpha({\bf
  16}) \ (\alpha = 1, 2, 3)$ are the three families of quark-lepton
multiplets.
The vacuum-expectation value of $\bar{\phi}$ of the GUT scale gives
the right-handed neutrinos Majorana masses of the order of $f_{\alpha
  \beta} \, v_\phi^2 / M_P \sim f_{\alpha \beta} \times 10^{13}$ GeV,
resulting in small masses for the left-handed neutrinos
\cite{lightN}.
Small masses of the quarks and leptons which generate sizable mixings
may be also induced by nonrenormalizable interactions \cite{mixing}.

It might appear that the non-vanishing U(1)$_A$ charges of $\phi$'s
would spoil the masslessness of the Higgs doublets $H_i$ since their
vacuum-expectation values break the U(1)$_A$ symmetry.
Fortunately enough, there remains an unbroken U(1) symmetry,
U(1)$_{A'}$, which is a linear combination of the original U(1)$_A$
and a U(1) subgroup of the SO(10)$_{GUT}$.
The global U(1)$_A$ and U(1)$_{A'}$ symmetries are physically
equivalent since the SO(10)$_{GUT}$ is a gauge symmetry.
Therefore, the Higgs $H_i$ are kept massless by the unbroken 
U(1)$_{A'}$ symmetry.

We conclude that the low-energy spectrum in the present vacuum is
nothing other than that in the SUSY standard model with the
right-handed neutrinos.

Now, several comments are in order.

First, the dangerous dimension-five operators \cite{dimfiveop} which
induce the proton decay are forbidden in our model by the U(1)$_{A'}$
symmetry.

Second, the SO(10)$_{GUT}$ is not asymptotically free above the
threshold of the GUT superfields such as $S_{IJ}$, $A_{IJ}$ and
$\phi$'s.
However, the gauge coupling constant $\alpha_{GUT}$ may not blow up
until the Planck scale even if all the GUT superfields have the GUT
scale masses.

Third, a SUSY-invariant mass term of the Higgs doublets may be
generated when the U(1)$_{A'}$ symmetry is broken spontaneously at an
intermediate scale \cite{murayama}.

Finally, we mention a possibility that the SO(10)$_{GUT}$ is broken
down to the SU(5)$_{GUT}$ at the Planck scale.
Then our model can be regarded as the SU(5)$_{GUT} \:\times$ SO(6)$_H$
theory up to the Planck scale.
That is, we may construct another model based on the SU(5)$_{GUT}
\:\times$ SO(6)$_H$ gauge group which also achieves natural
unification of the strong and electroweak interactions, as does the
SO(10)$_{GUT} \:\times$ SO(6)$_H$ model.


\begin{thebibliography}{99}
\bibitem{ours} There has been already an attempt for semisimple
  unification along the present line based on an SU(5)$_{GUT}
  \:\times$ SU(3)$_H$ gauge group \cite{our3,our4}.
  Although it retains phenomenological successes, it predicts
  deviation from the minimal SU(5) SUSY-GUT.
  See also Ref.\cite{our1,our2}, which deal with a U(1)$_H$ gauge
  group sacrificing the automatic hypercharge quantization.
\bibitem{anomaly} Of course, one may construct an anomaly-free model
  choosing real representations of SU($N_C$)$_H$ for the hyperquarks
  $Q_\alpha$.
\bibitem{vacuum} For $\lambda_Q v \neq m_Q$ or $-\frac{2}{3} m_Q$,
  there is no SUSY vacuum since the low-energy theory again has $N_f =
  1$.
\bibitem{anothervacuum} In the vacuum $\lambda_Q v = -\frac{2}{3}
  m_Q$, the massless hyperquarks $Q_\alpha^{11} \ (\alpha = 1, \cdots, 
    4)$ may be identified with a pair of Higgs doublets in the SUSY
  standard model.
  However, they do not have renormalizable Yukawa couplings to quarks
  and leptons \cite{our2}.
\bibitem{coupling} When the SO(6)$_C \:\times$ SU(2)$_L \:\times$
  SU(2)$_R$ is broken down to the standard SU(3)$_C \:\times$
  SU(2)$_L \:\times$ U(1)$_Y$ at the GUT scale, the gauge coupling
  constant of SU(3)$_C$ equals to $\alpha_6$ there.
\bibitem{chi} We introduce $\chi$ only to break the SO(10)$_{GUT}$
  down to the SU(5)$_{GUT}$.
  Instead of $\chi$, nonrenormalizable interactions such as
  $(\bar{\phi} \phi - \mu^2)^2$ serve for that purpose.
\end{thebibliography}
\end{document}